\documentclass[conference]{IEEEtran}
\IEEEoverridecommandlockouts

\usepackage{cite}
\usepackage{amsmath,amssymb,amsfonts}
\usepackage{algorithmic}
\usepackage{graphicx}
\usepackage{textcomp}
\usepackage{xcolor}
\usepackage{booktabs}
\usepackage{multirow}

\def\BibTeX{{\rm B\kern-.05em{\sc i\kern-.025em b}\kern-.08em
    T\kern-.1667em\lower.7ex\hbox{E}\kern-.125emX}}

\begin{document}

\setlength{\abovedisplayskip}{3pt}
\setlength{\belowdisplayskip}{3pt}
\setlength{\abovedisplayshortskip}{2pt}
\setlength{\belowdisplayshortskip}{2pt}

\title{Lightweight Time-Frequency Vibration Sensing for Consumer Device Condition Monitoring
\vspace{-0.3in}
\thanks{
This work was supported in part by the National Science and Technology Council (NSTC) of Taiwan under Grant 113-2926-I-001-502-G.
}
}

\author{
\IEEEauthorblockN{
Po-Heng Chou$^{1,3}$,
Jen-Yu Chiu$^{2}$,
Chun-Yu Yeh$^{2}$, Wei-Lung Mao$^{2}$, and Walid Saad$^{3}$
}
\IEEEauthorblockA{
$^{1}$Research Center for Information Technology Innovation (CITI), Academia Sinica (AS), Taipei 11529, Taiwan\\
$^{2}$Department of Electrical Engineering, National Yunlin University of Science and Technology (NYUST), Yunlin 64002, Taiwan\\
$^{3}$Bradley Department of Electrical and Computer Engineering (ECE), Virginia Tech (VT), Alexandria, VA 22305, USA\\
E-mails: d00942015@ntu.edu.tw, asd1248106@gmail.com, aa745579aa@gmail.com, wlmao@yuntech.edu.tw, walids@vt.edu
\vspace{-0.2in}
}
}

\maketitle


\begin{abstract}
Consumer rotating devices, including washing machines, air conditioners,
electric fans, robotic vacuum cleaners, and home fitness equipment, increasingly
rely on vibration sensing for intelligent condition monitoring and predictive
maintenance. Ensuring reliable operation in these resource-constrained appliances
requires efficient, lightweight artificial intelligence (AI) solutions capable of
processing non-stationary signals under real-world operational interference.
This paper proposes a lightweight AI-assisted vibration sensing framework that
employs continuous wavelet transform (CWT) and Hilbert-Huang transform (HHT)
to convert one-dimensional (1-D) vibration time-series signals into two-dimensional (2-D)
time-frequency representations. These representations are subsequently analyzed
using lightweight deep learning architectures, specifically MobileNetV3-Small, ConvNeXt-Tiny, and the You Only Look Once version 11 classification model (YOLOv11-cls). Public bearing benchmark datasets, including
Case Western Reserve University (CWRU), Paderborn University (PU), and Intelligent Maintenance Systems (IMS), are utilized as representative
rotating components to investigate model generalization. The proposed framework
is designed with eventual embedded deployment in consumer electronics in mind.
Experimental results demonstrate that CWT combined with YOLOv11-cls achieves
high diagnostic accuracy, while MobileNetV3-Small provides an attractive
low-latency alternative for computationally constrained edge platforms.
Conventional random splitting is also compared with dataset-specific structured isolation. The results highlight the importance of avoiding data leakage and quantify tradeoffs in classification accuracy, parameter count, floating-point operations (FLOPs), model footprint, preprocessing cost, and central processing unit (CPU) inference latency.
\end{abstract}

\begin{IEEEkeywords}
Vibration sensing, condition monitoring, consumer electronics, time-frequency analysis, lightweight deep learning, predictive maintenance.
\end{IEEEkeywords}


\section{Introduction}

Modern consumer rotating devices, including washing machines,
air conditioners, household electric fans, robotic vacuum cleaners,
and home fitness equipment, represent an important component of
contemporary smart home ecosystems. As consumers demand higher reliability,
energy efficiency, and quiet operation, intelligent condition monitoring
and early anomaly detection have become increasingly important for reducing
unexpected mechanical failures, maintenance requirements, and equipment
downtime. Vibration-based monitoring has been widely investigated for
mechanical fault diagnosis because rotating components such as motors,
bearings, shafts, and transmission mechanisms generate characteristic
dynamic responses during operation~\cite{zhang2020deep,peng2022survey}.

However, extracting reliable diagnostic indicators from vibration signals
presents substantial technical challenges. Vibration signals acquired from
rotating machinery are often non-stationary, highly transient, and affected
by changing operating conditions and environmental noise. Traditional
time-domain statistical indicators may not adequately capture weak or
incipient fault signatures. Conventional frequency-domain analysis, while
effective for identifying characteristic frequency components, does not
provide sufficient temporal localization of transient events. Consequently,
time-frequency representation (TFR) techniques have become important tools
for analyzing non-stationary vibration signals.

Among TFR techniques, the continuous wavelet transform (CWT) provides
multi-resolution analysis and simultaneous localization in time and scale~\cite{torrence1998practical}. The Hilbert-Huang transform (HHT), which
combines empirical mode decomposition (EMD) with Hilbert spectral analysis, provides an adaptive approach for nonlinear and non-stationary signals~\cite{huang1998empirical}. These approaches allow one-dimensional (1-D) vibration
signals to be represented as two-dimensional (2-D) time-frequency images, making
them suitable for subsequent convolutional neural network (CNN)-based
feature extraction~\cite{li2011bearing,wang2019deep}.

Recent work transformed vibration signals into CWT spectrograms and employed You Only Look Once (YOLO)-based detection to localize fault-related time-frequency regions,
thereby improving the observability and interpretability of vibration
fault characteristics~\cite{Chou2026CWTVibration}. In contrast, the
present study focuses on spectrogram-level classification rather than
time-frequency region detection.

The integration of artificial intelligence (AI) into consumer appliances
also requires models that achieve high diagnostic accuracy while satisfying
strict hardware constraints. Deep learning models deployed on embedded
microcontrollers, digital signal processors (DSPs), or low-power edge
system-on-chips (SoCs) must consider computational complexity, memory
footprint, and inference latency. Lightweight neural architectures have
therefore attracted considerable attention in edge-oriented applications.

Although extensive literature has investigated machine learning and deep
learning for mechanical fault diagnosis~\cite{zhang2020deep}, many studies
focus primarily on industrial rotating machinery. The computational
requirements and evaluation protocols of such studies do not necessarily
reflect the constraints of consumer electronic devices.

Unlike the localized time-frequency region detection framework in~\cite{Chou2026CWTVibration}, the present study focuses on lightweight
spectrogram-level classification, computational efficiency, and
generalization under dataset-specific structured isolation for potential
consumer device deployment.

The main contribution of this paper is to develop a lightweight AI-assisted vibration sensing framework for condition monitoring of consumer rotating devices. Although publicly available benchmark bearing datasets, namely Case Western Reserve University (CWRU), Paderborn University (PU), and Intelligent Maintenance Systems (IMS), are used as representative rotating components because
they provide established fault conditions and experimental configurations,
the underlying methodology is designed to be transferable to vibration
sensing modules embedded in consumer appliances.

The primary contributions of this work are summarized as follows:

\begin{itemize}
    \item We propose a lightweight AI-assisted vibration sensing framework
    conceptualized for intelligent condition monitoring in consumer rotating
    devices.

    \item We establish dataset-specific structured isolation strategies to
    reduce data leakage and provide a more realistic evaluation of model
    generalization across different benchmark data structures.

    \item We systematically compare CWT and HHT with respect to their ability
    to preserve transient vibration characteristics and their preprocessing
    computational costs.

    \item We evaluate lightweight deep learning architectures, including MobileNetV3-Small~\cite{howard2019searching}, YOLOv11-cls~\cite{khanam2024yolov11}, and ConvNeXt-Tiny~\cite{liu2022convnet}, and analyze the tradeoffs among diagnostic accuracy, parameter count, floating-point operations (FLOPs), model footprint, and central processing unit (CPU) inference latency.
\end{itemize}


\section{Related Work}

\subsection{Time-Frequency-Based Vibration Diagnosis}

To address the limitations of purely stationary frequency-domain analysis under fluctuating operating conditions, time-frequency methods have been widely adopted. CWT provides multi-resolution analysis via localized wavelet functions, effectively generating time-frequency images suitable for deep learning-based fault diagnosis~\cite{torrence1998practical,li2011bearing,wang2019deep}. Alternatively, HHT combines EMD with Hilbert spectral analysis for adaptive, data-driven decomposition~\cite{huang1998empirical}. While EMD extracts intrinsic mode functions (IMFs) without predefined basis functions, it often suffers from mode mixing and increased computational overhead when processing noisy signals.

\subsection{Lightweight Deep Learning for Condition Monitoring}

The development of lightweight neural networks has enabled deep learning
to move from cloud and high-performance computing platforms toward
resource-constrained edge devices. MobileNetV3 introduced efficient
depthwise separable convolution blocks and squeeze-and-excitation mechanisms
to achieve a favorable balance between computational cost and accuracy~\cite{howard2019searching}.

ConvNeXt subsequently demonstrated that modernized pure CNN architectures
can achieve strong visual representation performance while retaining the
advantages of convolutional networks~\cite{liu2022convnet}. ConvNeXt has
also been investigated for rotating component fault diagnosis and transfer
learning applications~\cite{xing2025fault}.

YOLO-based architectures have traditionally been developed for real-time
object detection. More recent versions also provide classification
configurations that can be applied to image-based diagnostic problems.
YOLOv11-cls is therefore considered in this study as a
high-performance alternative for classifying vibration spectrograms~\cite{khanam2024yolov11}.

It should be noted that programmable gradient information (PGI) is
associated with YOLOv9 rather than YOLOv11. YOLOv9 introduced PGI as a
training-related mechanism for improving information propagation~\cite{wang2024yolov9}. Therefore, PGI is not described as a YOLOv11
architectural component in this study.

Despite the development of lightweight architectures, evaluation procedures
remain an important concern. Randomly dividing highly correlated vibration
segments can cause samples originating from the same physical record to
appear in both training and testing sets. Such a procedure can result in
overly optimistic performance estimates. Recent benchmarking studies have
therefore emphasized more realistic splitting strategies that separate
physically distinct records or bearings between training and testing~\cite{hendriks2022benchmark,vieira2026realistic}.


\section{Proposed Framework and Methodology}

\begin{figure}[htbp]
\centerline{
\includegraphics[width=0.5\textwidth]{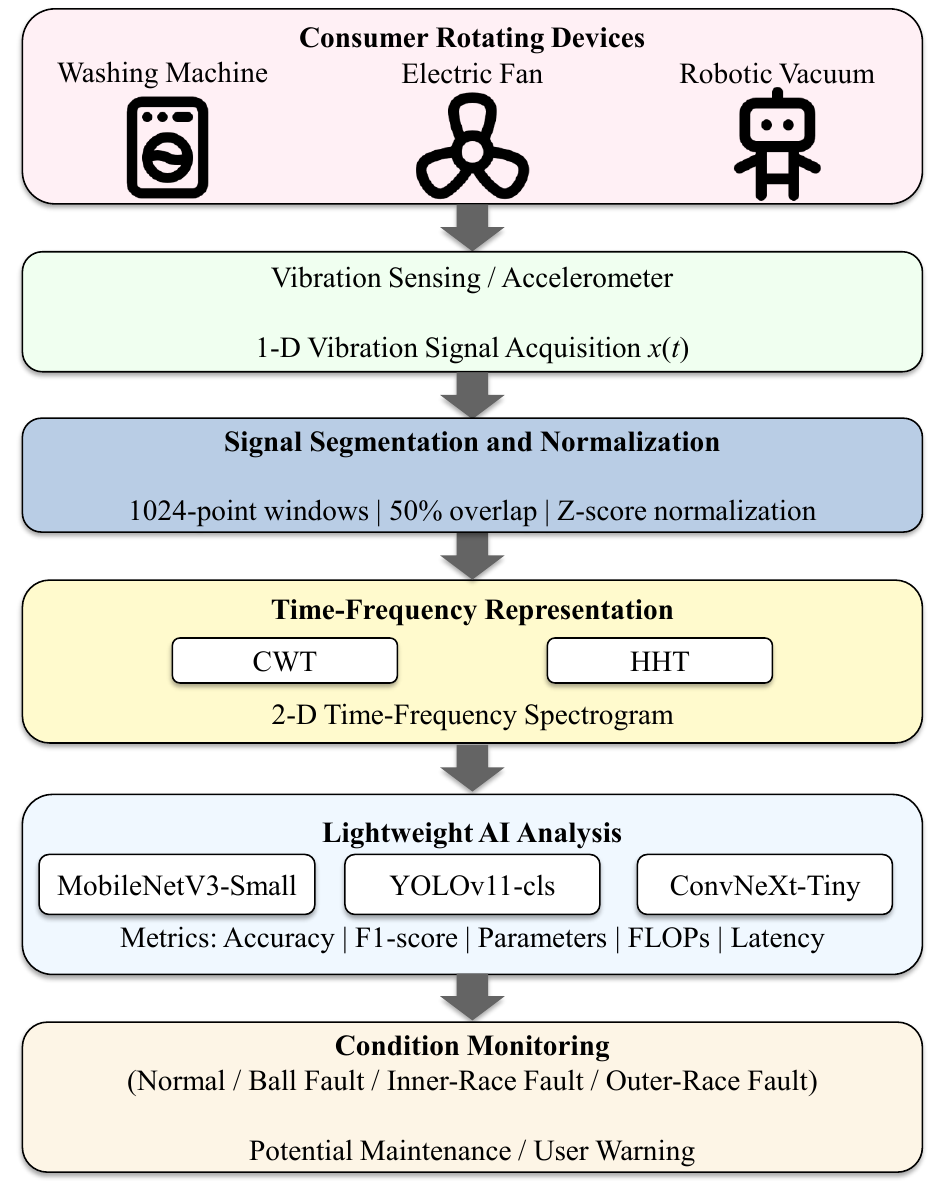}
}
\caption{Proposed framework comprising vibration acquisition, preprocessing, time-frequency representation, lightweight AI classification, and condition monitoring.}
\label{fig_arch}
\vspace{-0.1in}
\end{figure}

The proposed system architecture is designed as a potential application
framework for consumer electronics, as shown in Fig.~\ref{fig_arch}.
The workflow assumes that vibration signals are captured from a rotating
consumer device, such as a washing machine or household fan, using an
accelerometer. The raw vibration signal is first converted into a
time-frequency representation. The resulting representation is then
processed by a lightweight deep learning classifier, and the predicted
condition is finally provided to a condition monitoring system. Note that the condition monitoring encompasses four distinct categories: Normal, Ball Fault, Inner-Race Fault, and Outer-Race Fault.

\subsection{Time-Frequency Representation Techniques}

To capture non-stationary shock impulses and frequency-dependent vibration
characteristics, 1-D vibration signals $x(t)$ are converted
into 2-D time-frequency representations. Two transformation
methods are investigated: CWT and HHT.


\subsubsection{Continuous Wavelet Transform (CWT)}

The CWT provides localized analysis in both time
and scale. In this study, the Morlet wavelet is used as the mother wavelet.
The continuous wavelet transform is defined as
\begin{equation}
W_x(a,b) = \frac{1}{\sqrt{|a|}} \int_{-\infty}^{\infty} x(t) \psi^* \left( \frac{t-b}{a} \right) dt,
\end{equation}
where $W_x(a,b)$ is the CWT coefficient, $a$ is the scale, $b$ is the translation, $\psi(t)$ is the mother wavelet, and $^*$ denotes complex conjugation~\cite{torrence1998practical}. 

The transform provides localized information across multiple scales and preserves the temporal evolution of vibration energy. This representation is therefore suitable for CNN-based feature extraction.


\subsubsection{Hilbert-Huang Transform (HHT)}

The HHT combines empirical mode decomposition (EMD)
with Hilbert spectral analysis~\cite{huang1998empirical}. EMD adaptively
decomposes a vibration signal $x(t)$ into a finite number of intrinsic
mode functions (IMFs) and a residual component:
\begin{equation}
x(t) = \sum_{j=1}^{n} c_j(t) + r_n(t),
\end{equation}
where $j$ is the IMF index, $n$ is the number of IMFs, $c_j(t)$ is the $j$-th IMF, and $r_n(t)$ is the residual.

The Hilbert transform of each IMF is expressed as
\begin{equation}
\mathcal{H}[c_j(t)] = \frac{1}{\pi} P \int_{-\infty}^{\infty} \frac{c_j(\tau)}{t-\tau} d\tau,
\end{equation}
where $\mathcal{H}[\cdot]$ denotes the Hilbert transform, $P$ is the Cauchy principal value, and $\tau$ is the integration variable.

The corresponding analytic signal is defined as $z_j(t) = c_j(t) + i\mathcal{H}[c_j(t)] = a_j(t)e^{i\theta_j(t)}$, where $i$ is the imaginary unit, $a_j(t)$ is the instantaneous amplitude, and $\theta_j(t)$ is the instantaneous phase. The instantaneous angular frequency is then obtained directly as $\omega_j(t) = d\theta_j(t)/dt$. 

The Hilbert spectrum $H(\omega,t)$ can consequently be represented as
\begin{equation}
H(\omega,t) = \sum_{j=1}^{n} a_j(t) \delta \left[ \omega-\omega_j(t) \right],
\end{equation}
where $\omega$ denotes angular frequency and $\delta(\cdot)$ is the Dirac delta function. HHT provides adaptive decomposition and instantaneous-frequency information, which can be advantageous for transient vibration events. However, EMD-based processing is computationally iterative and can be sensitive to noise and mode mixing.

\begin{figure}[htbp]
\centering
\begin{minipage}{0.48\columnwidth}
  \centering
  \includegraphics[width=\linewidth]{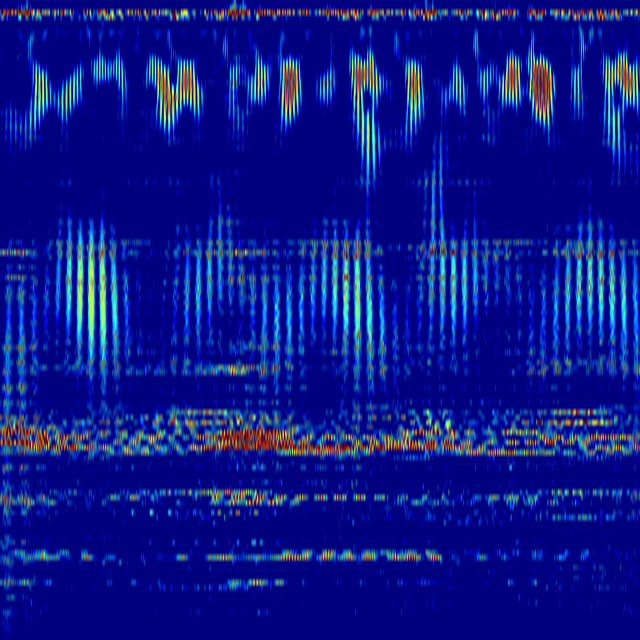}
  \caption{CWT spectrogram (CWRU fault).}
  \label{fig_cwt}
\end{minipage}\hfill
\begin{minipage}{0.48\columnwidth}
  \centering
  \includegraphics[width=\linewidth]{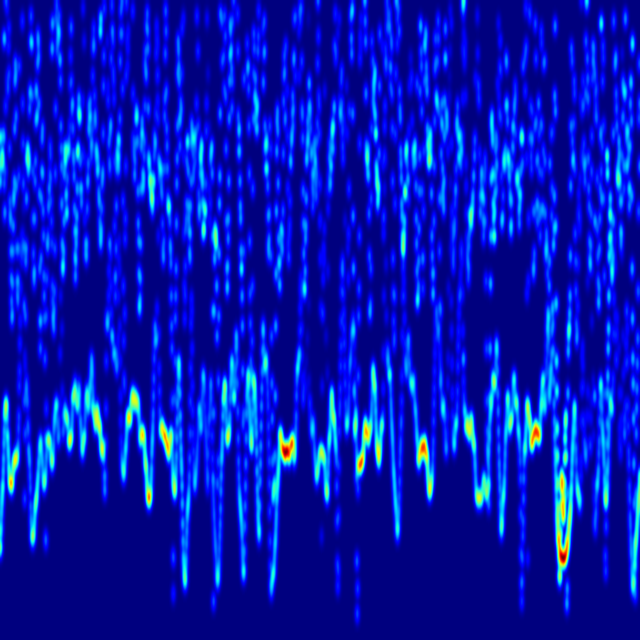}
  \caption{HHT spectrogram (CWRU fault).}
  \label{fig_hht}
\end{minipage}
\end{figure}

\vspace{-0.1in}
\subsection{Lightweight Deep Learning Architectures}

To satisfy potential embedded consumer-electronic constraints, three deep learning model families are evaluated:
1) \textbf{MobileNetV3-Small~\cite{howard2019searching}}: MobileNetV3 employs depthwise separable convolutions and lightweight squeeze-and-excitation mechanisms to achieve low computational complexity while maintaining effective feature extraction.
2) \textbf{ConvNeXt-Tiny~\cite{liu2022convnet}}: ConvNeXt modernizes conventional CNN architectures through design elements including large-kernel depthwise convolutions and other architectural improvements, providing strong visual feature extraction while retaining a convolution-based structure.
3) \textbf{YOLOv11-cls~\cite{khanam2024yolov11}}: YOLOv11-cls is the classification configuration of the YOLO11 family. It incorporates modern architectural components such as C3k2 blocks and C2 position-sensitive attention (C2PSA)-based feature processing. It is evaluated as a relatively higher-capacity classification model in comparison with MobileNetV3-Small.


\section{Experimental Setup and Evaluation}

\subsection{Datasets and Preprocessing}

To evaluate the proposed framework, three established bearing benchmark
datasets are utilized: the CWRU dataset,
the PU dataset, and the IMS bearing dataset.
These datasets provide representative rotating-component fault conditions
and have been widely used in machine-learning-based bearing diagnosis~\cite{neupane2020cwru,lessmeier2016pu,gousseau2016ims}.

The CWRU dataset provides vibration measurements under multiple motor-load
conditions and different bearing fault types. The PU dataset provides
multiple operating conditions and bearing damage categories, while the
IMS dataset provides run-to-failure vibration measurements that can be
used to investigate degradation-related characteristics.

\subsection{Implementation Details and Hyperparameters}

Signal preprocessing parameters were consistently defined for all
experiments. The 1-D vibration signals were segmented using
a sliding window of 1024 data points with a 50\% overlap ratio. Z-score
normalization was applied prior to time-frequency representation generation.

For CWT, the Morlet wavelet was used with scales ranging from 1 to 128.
For HHT, EMD was employed to extract the first three IMFs, which were empirically selected in this study to retain dominant vibration components while limiting the influence of high-frequency noise. A standard Gaussian filter was then
applied for visual smoothing. All generated 2-D spectrograms
were resized to $224\times224$ pixels to match the standard input
resolution of the selected CNN architectures.

Model training was conducted using the AdamW optimizer with an initial
learning rate of $1\times10^{-3}$, a batch size of 32, and 50 training
epochs. Transfer learning was performed using ImageNet-pretrained
weights. Hyperparameter optimization was conducted using Optuna with
20 search trials targeting learning rate and weight decay~\cite{akiba2019optuna}.

The software environment utilized PyTorch 2.1 and Python 3.9,
accelerated by an NVIDIA RTX 4090 graphics processing unit (GPU). CPU latency benchmarks were
evaluated using an Intel Core i7-13700K processor.


\subsection{Evaluation Protocols and Dataset-Specific Isolation Strategies}
Standard random splitting (e.g., 70\% training, 15\% validation, and 15\% testing) can introduce data leakage because adjacent overlapping segments may share highly similar signal characteristics and background noise profiles. To mitigate this issue, we adopted dataset-specific isolation strategies tailored to the inherent data structures of each benchmark. The generalized workflow is defined as: \textbf{Raw Data $\rightarrow$ Structured Isolation $\rightarrow$ Segmentation $\rightarrow$ CWT/HHT $\rightarrow$ Training/Testing}. The isolation unit is defined according to the characteristics of each dataset: load condition for CWRU, physical bearing identifier (ID) for PU, and chronological blocks for IMS. Segmentation is subsequently performed independently within each subset to prevent overlapping windows from the same designated isolation unit from being distributed across different subsets.

To ensure full reproducibility and explicitly demonstrate this rigorous methodology, Table~\ref{tab_detailed_split} outlines the isolation criteria and corresponding segment counts used across the three datasets. Because the data structures differ significantly, we applied a \textit{load-condition-wise split} for CWRU, a \textit{bearing-ID-wise split} for PU, and a \textit{chronological split} for IMS. In Table~\ref{tab_detailed_split}, HP denotes horsepower. For the PU dataset, H, IR, OR, and IROR denote healthy, inner-race fault, outer-race fault, and combined inner- and outer-race faults, respectively.

\begin{table*}[t]
\caption{Dataset Characteristics and Structured Isolation Settings}
\centering
\resizebox{0.85\textwidth}{!}{
\begin{tabular}{|l|c|l|l|l|c|}
\hline
\textbf{Dataset} & \textbf{Classes} & \textbf{Training Isolation} & \textbf{Validation Isolation} & \textbf{Testing Isolation} & \textbf{Train/Val/Test Segments} \\ \hline
\textbf{CWRU} & 4 & Loads: 1 HP, 2 HP & Load: 3 HP & Load: 0 HP & 6,174 / 1,265 / 1,265 \\ \hline
\textbf{PU} & 4 & \begin{tabular}[c]{@{}l@{}}H: K002-004\\ IR: KI01,05,08,16,18,21\\ OR: KA03-09,15,22\\ IROR: KB23,24,27\end{tabular} & \begin{tabular}[c]{@{}l@{}}H: K006\\ IR: KI03\\ OR: KA30\\ IROR: KI04\end{tabular} & \begin{tabular}[c]{@{}l@{}}H: K001\\ IR: KI07\\ OR: KA01\\ IROR: KI14\end{tabular} & 5,058 / 1,085 / 1,085 \\ \hline
\textbf{IMS} & 4 & Blocks: 0--84 per class & Blocks: 85--101 per class & Blocks: 102--118 per class & 340 / 68 / 68 \\ \hline
\end{tabular}
}
\vspace{-0.1in}
\label{tab_detailed_split}
\end{table*}

Specifically, for the CWRU dataset, the original sub-conditions are mapped into 4 fault categories (Normal, Ball Fault, Inner-Race Fault, Outer-Race Fault) encompassing multiple fault diameters (e.g., 0.007 to 0.021 in.). Because distinct bearing IDs are not consistently available across all conditions, a \textbf{load-condition-wise split} is utilized. For the PU dataset, under the N09\_M07\_F10 operating condition, a \textbf{bearing-ID-wise split} allocates mutually exclusive physical bearing IDs to the respective subsets to ensure strict structural isolation. For the IMS dataset, four long-duration time-frequency representations were generated separately for the Normal, Ball Fault, Inner-Race Fault, and Outer-Race Fault classes. Each class-specific representation was then divided into sequential chronological blocks. For every class, blocks 0–84 were assigned to training, blocks 85–101 to validation, and blocks 102–118 to testing. Window segmentation was performed independently within each subset. This procedure prevents temporally adjacent and overlapping windows from being distributed across different subsets. The resulting training, validation, and testing sets contain 340, 68, and 68 segments, respectively, with equal contributions from the four fault classes.

\vspace{-0.05in}
\section{Experimental Results and Discussion}


\subsection{Preprocessing Latency: CWT vs. HHT}
Since preprocessing latency contributes directly to the total response time of an embedded condition-monitoring system, the CPU execution time required to generate a single $224\times224$ time-frequency representation was evaluated.
To systematically quantify this preprocessing overhead, we measured the latency over a minimum of 100 independent runs on an Intel Core i7-13700K CPU with a batch size of one, with all results reported as mean $\pm$ standard deviation. The CWT preprocessing required only $\mathbf{15.2 \pm 1.8}$ \textbf{ms} per sample. In stark contrast, the iterative sifting process associated with EMD caused HHT preprocessing to require $\mathbf{88.5 \pm 7.4}$ \textbf{ms} per sample. 

This confirms that HHT incurs an approximately six-fold latency overhead compared to CWT. Although HHT provides adaptive instantaneous-frequency analysis, its substantially higher preprocessing cost makes CWT a much more attractive candidate for real-time consumer-device applications under the tested configuration.


\subsection{Classification Performance Comparison: CWT vs. HHT}
To provide a controlled comparison between CWT and HHT, the MobileNetV3-Small architecture, dataset-specific isolation settings, training hyperparameters, and evaluation procedures were kept identical for both representations. Table~\ref{tab_cwt_hht_compare} presents the accuracy, precision, recall, and F1-score obtained using CWT and HHT across the three benchmark datasets.

\begin{table}[htbp]
\caption{Classification Performance of CWT and HHT under Dataset-Specific Isolation (MobileNetV3-Small)}
\centering
\resizebox{\columnwidth}{!}{
\begin{tabular}{|l|c|c|c|c|c|}
\hline
\textbf{Dataset} & \textbf{TFR} & \textbf{Accuracy (\%)} & \textbf{Precision (\%)} & \textbf{Recall (\%)} & \textbf{F1-score (\%)} \\ \hline
\multirow{2}{*}{CWRU} & CWT & 92.1 & 91.8 & 92.1 & 91.9 \\ \cline{2-6}
 & HHT & 89.5 & 89.1 & 89.5 & 89.3 \\ \hline
\multirow{2}{*}{PU} & CWT & 86.3 & 85.9 & 86.3 & 86.0 \\ \cline{2-6}
 & HHT & 81.2 & 80.5 & 81.2 & 80.8 \\ \hline
\multirow{2}{*}{IMS} & CWT & 89.2 & 88.7 & 89.2 & 88.9 \\ \cline{2-6}
 & HHT & 85.4 & 84.9 & 85.4 & 85.1 \\ \hline
\end{tabular}
}
\vspace{-0.1in}
\label{tab_cwt_hht_compare}
\end{table}

Based on the classification results in Table~\ref{tab_cwt_hht_compare} and the preprocessing latency presented above, CWT consistently outperforms HHT across all three evaluated datasets. HHT provides adaptive instantaneous-frequency analysis but is susceptible to mode mixing in noisy environments, which may contribute to less stable CNN feature extraction. In addition, HHT requires approximately six times the preprocessing latency of CWT. These results indicate that CWT provides a more favorable accuracy-latency tradeoff for resource-constrained consumer devices under the tested configuration.


\begin{table*}[t]
\caption{Classification Performance and Generalization Under Random and Dataset-Specific Structured Isolation (Evaluated using CWT)}
\centering
\resizebox{0.9\textwidth}{!}{%
\begin{tabular}{|l|c|c||c|c|c|c||c|c|c|c|c|}
\hline

\multirow{2}{*}{\textbf{Model Architecture}}
&
\multirow{2}{*}{\textbf{Dataset}}
&
\multirow{2}{*}{\textbf{TFR}}
&
\multicolumn{4}{c||}{\textbf{Random Split (\%)}}
&
\multicolumn{5}{c|}{\textbf{Structured Isolation (\%)}} 
\\
\cline{4-12}

&
&
&
\textbf{Accuracy}
&
\textbf{Precision}
&
\textbf{Recall}
&
\textbf{F1}
&
\textbf{Accuracy}
&
\textbf{Precision}
&
\textbf{Recall}
&
\textbf{F1}
&
\textbf{Drop}
\\
\hline

\multirow{3}{*}{MobileNetV3-Small~\cite{howard2019searching}}
& CWRU
& CWT
& 99.8
& 99.7
& 99.8
& 99.7
& 92.1
& 91.8
& 92.1
& 91.9
& -7.7
\\
\cline{2-12}

& PU
& CWT
& 97.4
& 97.2
& 97.4
& 97.3
& 86.3
& 85.9
& 86.3
& 86.0
& -11.1
\\
\cline{2-12}

& IMS
& CWT
& 98.3
& 98.1
& 98.3
& 98.2
& 89.2
& 88.7
& 89.2
& 88.9
& -9.1
\\
\hline

\multirow{3}{*}{YOLOv11-cls~\cite{khanam2024yolov11}}
& CWRU
& CWT
& 100.0
& 100.0
& 100.0
& 100.0
& 96.5
& 96.3
& 96.5
& 96.4
& -3.5
\\
\cline{2-12}

& PU
& CWT
& 99.5
& 99.4
& 99.5
& 99.4
& 90.1
& 89.8
& 90.1
& 89.9
& -9.4
\\
\cline{2-12}

& IMS
& CWT
& 100.0
& 100.0
& 100.0
& 100.0
& 93.2
& 93.1
& 93.2
& 93.1
& -6.8
\\
\hline

\multirow{3}{*}{ConvNeXt-Tiny~\cite{liu2022convnet}}
& CWRU
& CWT
& 99.9
& 99.9
& 99.9
& 99.9
& 94.8
& 94.5
& 94.8
& 94.6
& -5.1
\\
\cline{2-12}

& PU
& CWT
& 98.2
& 98.0
& 98.2
& 98.1
& 88.0
& 87.6
& 88.0
& 87.8
& -10.2
\\
\cline{2-12}

& IMS
& CWT
& 99.5
& 99.4
& 99.5
& 99.4
& 91.4
& 91.0
& 91.4
& 91.2
& -8.1
\\
\hline

\end{tabular}%
}
\vspace{-0.1in}
\label{tab_performance}
\end{table*}

\subsection{Classification Performance and Generalization}

Table~\ref{tab_performance} compares the classification performance of MobileNetV3-Small~\cite{howard2019searching}, YOLOv11-cls~\cite{khanam2024yolov11}, and ConvNeXt-Tiny~\cite{liu2022convnet} under random and dataset-specific structured isolation. Across
all evaluated architectures and datasets, dataset-specific structured isolation produces
lower accuracy than conventional random splitting.

The observed accuracy reductions range from 3.5 percentage points to
11.1 percentage points. This difference indicates that random splitting
can produce overly optimistic results when highly correlated vibration
segments are distributed across the training and testing subsets.

Among the evaluated architectures, YOLOv11-cls demonstrates the strongest
performance under the dataset-specific structured isolation protocols, achieving 96.5\% accuracy on the CWRU dataset.
The corresponding MobileNetV3-Small accuracy is 92.1\%. Although
MobileNetV3-Small sacrifices some diagnostic accuracy compared with
YOLOv11-cls, its substantially lower computational complexity makes it
attractive for highly constrained edge platforms.

ConvNeXt-Tiny provides intermediate performance between MobileNetV3-Small
and YOLOv11-cls on the CWRU dataset, while requiring a substantially
larger model footprint.


\subsection{Complexity and Accuracy-Efficiency Tradeoffs}
To ensure fair and reproducible complexity comparisons presented in Table~\ref{tab_complexity}, all models were evaluated using a fixed input resolution of $224 \times 224$. The parameters and FLOPs were quantified using the THOP (version 0.1.1) profiling tool. Since the checkpoint formats may differ across frameworks, the measured checkpoint file sizes are treated as implementation-specific storage measurements rather than strict architecture-level memory requirements. For the CPU latency tests, measurements were rigorously performed on an Intel Core i7-13700K processor with all GPU execution paths disabled (CPU only) and the batch size set to 1. An initial 100-iteration warm-up phase was executed, followed by at least 1,000 formal inference runs. The measured network inference latency includes only neural network execution and excludes data loading and CWT/HHT preprocessing.


\begin{table}[htbp]
\caption{Computational Complexity and Network CPU Inference Latency}
\centering
\setlength{\tabcolsep}{2pt}
\resizebox{\columnwidth}{!}{%
\begin{tabular}{|l|c|c|c|c|}
\hline
\textbf{Model} &
\textbf{Parameters (M)} &
\textbf{FLOPs (G)} &
\begin{tabular}[c]{@{}c@{}}\textbf{Checkpoint}\\\textbf{Size (MB)}\end{tabular} &
\textbf{Latency (ms)} \\ \hline
MobileNetV3-Small~\cite{howard2019searching} & 2.54 & 0.06 & 9.8 & $8.3 \pm 0.6$ \\ \hline
YOLOv11-cls~\cite{khanam2024yolov11} & 10.35 & 2.43 & 19.9 & $25.5 \pm 2.5$ \\ \hline
ConvNeXt-Tiny~\cite{liu2022convnet} & 27.80 & 4.46 & 109.1 & $28.9 \pm 1.5$ \\ \hline
\end{tabular}%
}
\vspace{-0.1in}
\label{tab_complexity}
\end{table}

As shown in Table~\ref{tab_complexity}, model architecture strongly
affects the feasibility of embedded deployment.
MobileNetV3-Small has a notably low parameter count of 2.54 M and a computational cost of only 0.06 GFLOPs.
YOLOv11-cls requires higher computational resources (2.43 GFLOPs) than MobileNetV3-Small but
remains significantly more efficient than ConvNeXt-Tiny. Furthermore, its
classification accuracy under structured isolation is consistently the highest across the evaluated datasets.
ConvNeXt-Tiny has the largest parameter count (27.80 M) and model footprint (109.1 MB) among
the three evaluated architectures. Its measured CPU latency is also
the highest.

Considering both preprocessing and model inference, MobileNetV3-Small provides a strong candidate for severely constrained edge platforms, while YOLOv11-cls is more appropriate when higher diagnostic accuracy is prioritized. The total diagnostic latency for a single vibration sample in an end-to-end scenario is the sum of the TFR preprocessing latency and the network inference latency. For example, utilizing MobileNetV3-Small combined with CWT preprocessing yields a total processing time of approximately \textbf{23.5 ms} ($15.2 + 8.3$). This result demonstrates the potential of combining efficient time-frequency preprocessing with lightweight CNN inference for near-real-time condition monitoring.


\section{Potential Applications in Consumer Electronics}

The proposed framework is designed to be transferable to vibration
sensing modules embedded in various consumer rotating devices.

Potential applications include: 1) \textbf{Smart Washing Machines}: detecting drum imbalance, motor bearing degradation, and suspension wear during high-speed spin cycles; 2) \textbf{Household Electric Fans}: identifying early-stage sleeve or ball-bearing wear and abnormal rotating behavior; and 3) \textbf{Robotic Vacuum Cleaners}: diagnosing drive-wheel motor anomalies, main-brush gearbox problems, and mechanical interference caused by accumulated debris. 


\vspace{-0.05in}

\section{Limitations and Future Work}
The standardized datasets provide reproducible fault conditions, but they do not fully represent the acoustic, mechanical, and environmental complexity of domestic appliances. Differences in device structures, sensor positions, and operating loads may also affect cross-device generalization. Moreover, the CPU latency results provide only a hardware-specific baseline and do not directly represent execution time on resource-constrained embedded microcontrollers.

Future work will collect consumer-appliance vibration data and investigate cross-device generalization and hardware-aware optimization, including 8-bit integer (INT8) quantization, pruning, and knowledge distillation.


\section{Conclusion}

In this paper, we have developed a lightweight AI-assisted vibration sensing framework for consumer rotating devices. The results show that CWT provides a more favorable accuracy-latency tradeoff than HHT. Under dataset-specific structured isolation, YOLOv11-cls achieves the highest accuracy, whereas MobileNetV3-Small offers greater efficiency for constrained platforms. The performance gap between random and structured splitting demonstrates the importance of preventing correlated segments from entering both training and testing sets. These results provide a foundation for future deployment in consumer devices.


\bibliographystyle{IEEEtran}

\end{document}